\documentclass[lettersize,journal]{IEEEtran}
\pdfoutput=1
\usepackage{amsmath,amsfonts}
\usepackage{algorithmic}
\usepackage{algorithm}
\usepackage{array}
\usepackage[caption=false,font=normalsize,labelfont=sf,textfont=sf]{subfig}
\usepackage{textcomp}
\usepackage{stfloats}
\usepackage{url}
\usepackage{verbatim}
\usepackage{graphicx}
\usepackage{cite}
\usepackage{amssymb}
\newcommand{\RNum}[1]{\uppercase\expandafter{\romannumeral #1\relax}}
\begin{document}

\title{Beam Squint Assisted User Localization in Near-Field Communications Systems}
\author{Hongliang Luo and Feifei Gao
\thanks{H. Luo and F. Gao are with Institute for Artificial Intelligence, Tsinghua University (THUAI),
State Key Lab of Intelligent Technologies and Systems, Tsinghua University,
Beijing National Research Center for Information Science and Technology (BNRist),
Department of Automation, Tsinghua University
Beijing, P.R. China (email: 19020100142@stu.xidian.edu.cn, feifeigao@ieee.org).}
}



\maketitle

\begin{abstract}
The beam squint phenomenon in  massive multi-input and multi-output wideband communications has been widely concerned recently, which generally deteriorates the beamforming performance. In this paper, we find that with the aid of the time-delay lines (TDs), the range and trajectory of the beam squint of a near-field  communications system can be freely controlled, and hence it is possible to reversely utilize the beam squint for user localization.
We derive the trajectory equation for \emph{near-field beam squint points} and design a way to control the trajectory of these beam squint points. With the proposed design, beamforming from different subcarriers would purposely point to different angles and different distances such that users from different positions would receive the maximum power at different subcarriers. Hence, one can simply find the different users' position from the beam squint effect.  Simulation results demonstrate the effectiveness of the proposed scheme.
\end{abstract}


\section{Introduction}

The sixth generation of mobile communications (6G) will adopt massive multiple input and multiple output (MIMO) and higher frequency bands, e.g., mmWave communications or  Terahertz communication. In this case, the phenomenon of \emph{beam squint} would appear, in which the beamforming from at different subcarriers would point to different direction, making the energy of part subcarriers deviate from the user position \cite{a1,a2}. Beam squint effect is usually considered as a negative effect and  should be mitigated with the aid of the time-delay lines (TDs) \cite{b1,b2,b3}. When the user moves to the BS, say less than Rayleigh distance, the communication electromagnetic field would change and the near-field effect would appear. 
With near field assumption, wideband systems will also occur beam squint phenomenon, while different subcarriers will focus on different positions \cite{c1}.

On the other side, integrated sensing and communications (ISAC) has attracted much research interest recently and is deemed as one of the key technologies of 6G \cite{d1}. The idea is to use the communication signaling to sense the user's position as a simultaneous functionality. Once the user's position is obtained, the base station system can  better serve the user's communication \cite{d2}.  The sensing can also be divided into far-field sensing and near-field sensing, depending on the distance between BS and the user. With a single BS, far-field sensing could only obtain user's angle. Nevertheless, it is possible to sense user's angle and distance, i.e., the users' position for near-field sensing. 
J. Yang proposed a localization method based on extremely large lens antenna array, which utilizes window effect for energy focusing property of ExLens \cite{yang1}.
Y. Lin proposed a Twin-IRS-assisted user localization method, which achieves high-precision user localization \cite{L1}.
However, these methods do not consider the effects of wideband systems and requires large times wave sweeps.

In this paper, we propose a fast near field user localization algorithm by reversely utilize the beam squint effect. The contributions of this paper are summarized as follows.
\begin{itemize}
\item We analyze the near-field beam squint phenomenon and derive the trajectory equation of \emph{near-field beam squint points}.

\item We reversely utilize the beam squint for user localization and design a way to control the range and trajectory of the beam squint points.

\item We propose a fast near-field user position sensing scheme based on TDs-assisted beam squint, and the simulation results demonstrate the effectiveness of the proposed scheme.
\end{itemize}

The remainder of the paper is organized as follows. In Section \RNum{2}, we introduce the system model of the near-field communications.
In Section \RNum{3}, the near-field beam squint phenomenon is analyzed and the trajectory equation for the squint points is derived. Then we propose a near-field user positioning scheme in Section \RNum{4}. Simulation results and conclusion are given in Section \RNum{5} and Section \RNum{6}.

\section{System Model}
We consider a wideband massive MIMO system with orthogonal frequency division multiplexing (OFDM) modulation.
The base station (BS) is equipped with an $N$-antenna uniform linear array (ULA) with antenna spacing $d$ and a single RF chain. 
All antennas are deployed on the $y$-axis at position (0, $nd$), $n=-\frac{N-1}{2}$, ...,$\frac{N-1}{2}$,
and the antenna array aperture is $D=(N-1)d\approx Nd$. 
The carrier frequency and transmission bandwidth are denoted as ${f}_c$ and $W$,
and then the frequency range is $[{f}_c-\frac{W}{2},{f}_c+\frac{W}{2}]$.
Assuming there are total $M+1$ subcarriers, 
the $0$-th subcarrier has the lowest frequency $f_0={f}_c-\frac{W}{2}$
and then the $m$-th subcarrier frequency is $f_m={f}_0+m\frac{W}{M}$.
Particularly, let us denote the baseband frequency of the $m$-th subcarrier as
$\widetilde{f}_m=m\frac{W}{M}$, where $m=0,1,2,...,M$.
Clearly, $f_m=f_0+\widetilde{f}_m$.
So the $M$-th subcarrier has the highest frequency, the passband frequency is 
$f_M=f_0+W$, and the baseband frequency is $\widetilde{f}_M=W$.

Assume that the BS serves $K$ users, and each user is equipped with a single antenna.
Suppose that the $k$-th user is located at $(x_k,y_k)$, and corresponding polar coordinate is $(r_k,\theta_k)$.
Then the distance between the $n$-th antenna and the $k$-th user is $r_{k,n}=\sqrt{x_k^2+(y_k-nd)^2}$.

According to \cite{ruili}, the boundary between the near-field and far-field is determined by the Rayleigh distance $Z=\frac{2D^2}{\lambda}$.
Hence, all users are assumed to be in the near-field range.
Under the assumption of near-field, we consider only one line-of-sight (LoS) channel between the user and the BS while the proposed study can be readily extended to the multipath scenario. 
The LoS channel $\mathbf{h}(x_k,y_k,{f}_m)\in \mathbb{C}^{N\times 1}$ is modeled as \cite{xindao}

\begin{figure}
\centering
\includegraphics[width=2.5in]{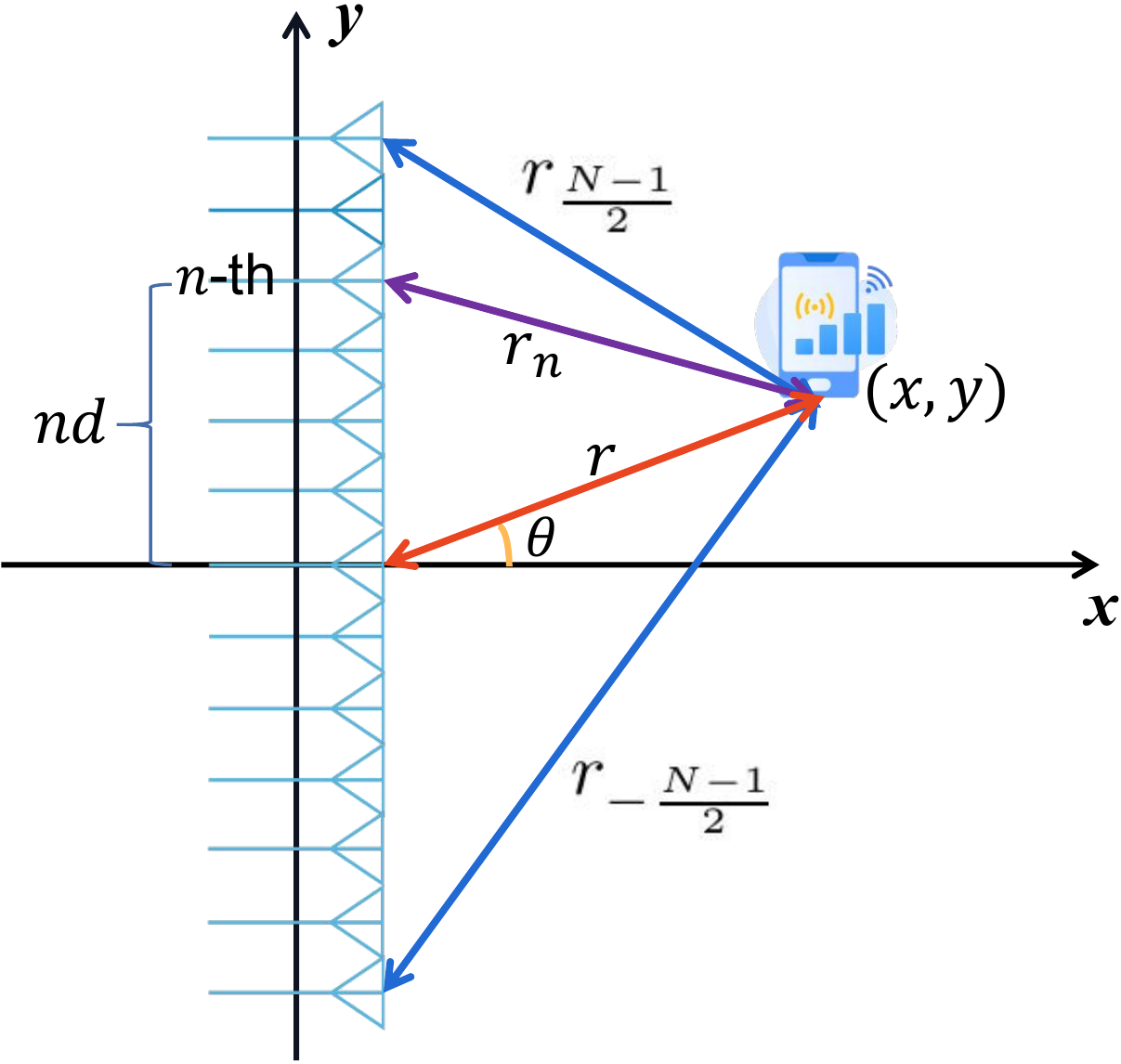}
\caption{Near-field system model.}
\label{fig1:env}
\end{figure}

\begin{equation}
\label{deqn_ex1a}
[\mathbf{h}(x_k,y_k,{f}_m)]_n=\sqrt{\beta(f_m)}\frac{e^{-j2\pi f_m\frac{r_{k,n}}{c}}}{r_{k,n}}
\end{equation}
where $[\mathbf{h}(x_k,y_k,{f}_m)]_n$ denotes the channel between the $n$-th antenna and the $k$-th user, and
$\frac{\beta(f_m)}{r_{k,n}^2}=(\frac{c}{4\pi f_mr_{k,n}})^2$ denotes the free-space path loss.

In order to provide better communications link for users, BS adopt the phase-shifters (PSs) based beamforming for the single RF chain.
We consider that the optimal beamforming vector of near-field $\mathbf{w}(x_0,y_0)\in \mathbb{C}^{N\times 1}$ is determined by the point $(x_0,y_0)$ and the lowest carrier frequency $f_0$.
Then $\mathbf{w}(x_0,y_0)$ should match the phase of $[\mathbf{h}(x_0,y_0,f_0)]$ and is denoted as

\begin{equation}
\label{deqn_ex1a}
[\mathbf{w}(x_0,y_0)]_n=\frac{1}{\sqrt{N}}\arg([\mathbf{h}(x_0,y_0,f_0)]_n)=\frac{1}{\sqrt{N}}e^{-j2\pi f_0\frac{r_{0,n}}{c}}
\end{equation}
where $\arg(x)$ denotes the phase of $x$. It is worth pointing out that phase $2\pi f_m\frac{r_{k,n}}{c}$ is nonlinear to $n$ in (1) and (2).

\section{Near-Field Beam Squint}

With a single RF chain, the beamforming vectors on all subcarriers would be the same $\mathbf{w}(x_0,y_0)$, which is the case for narrowband communications.
With wideband communications, i.e., $W$ is large, the beamforming of the $m$-th subcarrier may not point to the target user position $(x_0,y_0)$. Such a phenomenon is named as beam squint as the beamforming direction gradually ``squint'' over the frequency.  When $\mathbf{w}(x_0,y_0)$ points to the position $(x_0,y_0)$, we can compute the power received at any position $(x, y)$ on the $m$-th subcarrier as 

\begin{equation}
\begin{split}
\begin{aligned}
\label{deqn_ex1a}
&g(x,y,{f}_m,x_0,y_0)=\left| \mathbf{h}^H(x,y,{f}_m)\cdot\mathbf{w}(x_0,y_0)\right|\\
&=\sqrt{\frac{\beta({f}_m)}{N}}\left| \sum _{n=-\frac{N-1}{2}}^{\frac{N-1}{2}} 
\frac{1}{r_n}e^{j2\pi f_m\frac{r_n}{c}}e^{-j2\pi f_0\frac{r_{0,n}}{c}}
\right|
\end{aligned}
\end{split}
\end{equation}
where $r_n=\sqrt{x^2+(y-nd)^2}$ denotes the distance between the near-field point $(x,y)$ and the $n$-th BS antenna.


From (3) we know the received power $g(x,y,{f}_m,x_0,y_0)$ is inversely proportional to $r_{n}$, but in communication systems, we are more concerned with phase factors.
When $m=0$, ${g}$ is maximized at the position $(x,y)=(x_0,y_0)$.
However, for the other subcarriers $f_m (m\neq0)$, the position
$(x,y)$ that maximize ${g}$ is not $(x_0,y_0)$, and is defined as $(x_m,y_m)$,
i.e., the beamforming focus squint to other position. 
We call $(x_m,y_m)$ as \emph{near-field beam squint points}, 
and its polar coordinate is $(r_m,\theta_m)$.
Each $(x_m,y_m)$ should satisfy the following equations

\begin{equation}
\label{deqn_ex1a}
f_m\frac{r_{m,n}}{c}- f_0\frac{r_{0,n}}{c}=p_n
\end{equation}
for $n=-\frac{N-1}{2}$, ...,$\frac{N-1}{2}$ and $m=0,1,2,...,M$,
where $p_n\in\mathbb{Z}$ is an integer introduced to align the phase of all the additive terms.

In order to derive $(r_m,\theta_m)$ the equation (4),
we use the Fresnel approximation \cite{jinsi}
for $r_n=\sqrt{x^2+(y-nd)^2}$ as

\begin{equation}
\begin{split}
\begin{aligned}
\label{deqn_ex1a}
r_n&=\sqrt{x^2+(y-nd)^2}=\sqrt{(r\cos\theta)^2+(r\sin\theta-nd)^2}\\
&=\sqrt{r^2+n^2d^2-2rnd\sin\theta}=r\sqrt{1+(\frac{n^2d^2}{r^2}-\frac{2nd\sin\theta}{r})}\\
&\approx r[1+(\frac{n^2d^2}{r^2}-\frac{2nd\sin\theta}{r})-\frac{1}{8}(\frac{n^2d^2}{r^2}-\frac{2nd\sin\theta}{r})^2]\\
&=r-nd\sin\theta+\frac{n^2d^2\cos^2\theta}{2r}.
\end{aligned}
\end{split}
\end{equation}

Next we apply
the Fresnel approximation for both the center point $r_{0,n}$ and beam squint point $r_{m,n}$ in equation (4) and obtain
\begin{equation}
\begin{split}
\begin{aligned}
\label{deqn_ex1a}
\left[\frac{f_m}{c}(r_m+\frac{n^2d^2\cos^2\theta_m}{2r_m})
-\frac{f_0}{c}(r_0+\frac{n^2d^2\cos^2\theta_0}{2r_0})\right]\\
-\left[\frac{f_m}{c}nd\sin\theta_m-\frac{f_0}{c}nd\sin\theta_0\right]=p_n.
\end{aligned}
\end{split}
\end{equation}

Because the latter term $\frac{f_m}{c}nd\sin\theta_m-\frac{f_0}{c}nd\sin\theta_0$ of (6) does not explicit contain $r_m$ and $r_0$, and because the equation (6) should be independent of $n$, we can
let $\frac{f_m}{c}nd\sin\theta_m-\frac{f_0}{c}nd\sin\theta_0=0$.
Then we obtain the angle constraint equation for near-field beam squint points as
\begin{equation}
\begin{split}
\begin{aligned}
\label{deqn_ex1a}
\sin\theta_m=\frac{f_0}{f_m}\sin\theta_0.
\end{aligned}
\end{split}
\end{equation}

\begin{figure}[H]
\centering
\includegraphics[width=3in]{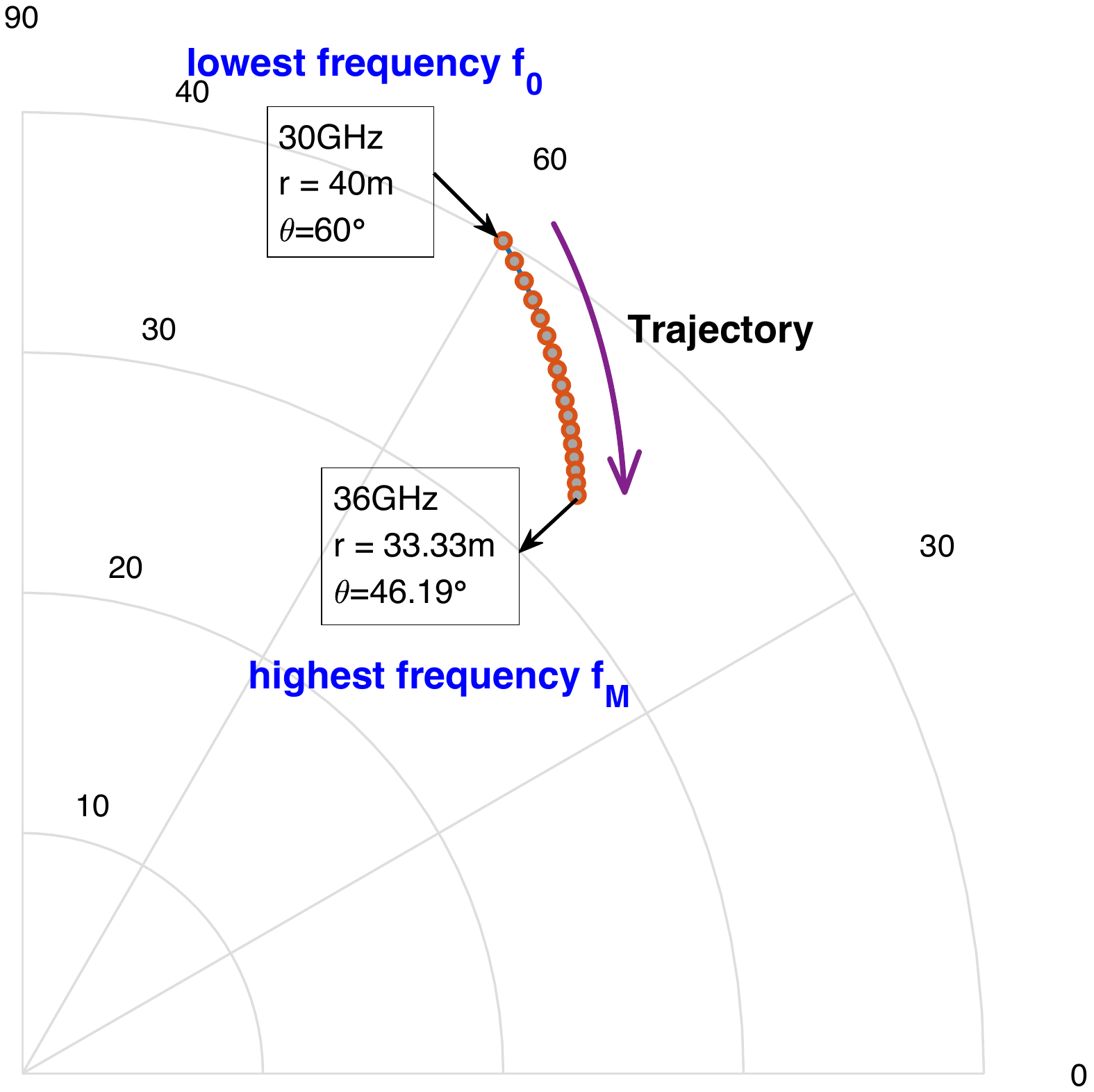}
\caption{Trajectory of beam squint points. Where $f_0=$ 30 GHz, $W=$ 6 GHz, $M=$16, the lowest frequency carrier $f_0$ is set to focus on $(40m, 60^{\circ}).$}
\label{fig_1}
\end{figure}

Then, the equation (6) reduces to
\begin{equation}
\begin{split}
\begin{aligned}
\label{deqn_ex1a}
\frac{f_m}{c}\left(r_m+\frac{n^2d^2\cos^2\theta_m}{2r_m}\right)
-\frac{f_0}{c}\left(r_0+\frac{n^2d^2\cos^2\theta_0}{2r_0}\right)=p_n.
\end{aligned}
\end{split}
\end{equation}

For $n=0$, we obtain from (8) that
\begin{equation}
\begin{split}
\begin{aligned}
\label{deqn_ex1a}
\frac{f_m}{c}r_m-\frac{f_0}{c}r_0=p_0.
\end{aligned}
\end{split}
\end{equation}

We then bring (9) into (8) and obtain
\begin{equation}
\begin{split}
\begin{aligned}
\label{deqn_ex1a}
\frac{d^2}{2c}\left(\frac{f_m\cos^2\theta_m}{r_m}-\frac{f_0\cos^2\theta_0}{r_0}\right)=\frac{p_n-p_0}{n^2}.
\end{aligned}
\end{split}
\end{equation}

Since (10) holds for arbitrary $n$, we know that
$\frac{f_m\cos^2\theta_m}{r_m}-\frac{f_0\cos^2\theta_0}{r_0}=0$. 
Then we obtain the distance constraint equation for near-field beam squint points as
\begin{equation}
\begin{split}
\begin{aligned}
\label{deqn_ex1a}
r_m=r_0\cdot\frac{f_m}{f_0}\cdot\frac{\cos^2\theta_m}{\cos^2\theta_0}
\end{aligned}
\end{split}
\end{equation}
where $\cos^2\theta_m=1-\sin^2\theta_m=1-(\frac{f_0}{f_m}\sin\theta_0)^2$.

In summary, the angle and distance of the beam squint point $(x_m,y_m)$ of the $m$-th sub-carrier is determined by formula (7) and (11), respectively.

We take an example in Fig. 2, where it is see that
as the subcarrier frequency $f_m$ increases, beamforming from different sybcarriers would focus on different positions and can be connected into a curve trajectory.
The start point of the trajectory is determined by the $0$-th subcarrier ${f}_0$ while the end point of the trajectory is determined by the $M$-th subcarrier ${f}_M$.
As the subcarrier frequency increases, both the $r_m$ and $\sin\theta_m$  will decrease.
In Fig. 2, the lowest frequency is $f_0=30$ GHz whose beamforming focuses at 
$(40m, 60^{\circ})$, while the highest frequency subcarrier is $36$ GHz whose beamforming squint to $(33.33m, 46.19^{\circ})$.
Clearly, beam squint cannot be ignored in the wideband system.

Interestingly, the trajectory of beam squint allows BS to distinguish users at different position based on the power received at different subcarriers, which allows us to reduce the number of beam sweeping over the time. 
By doing this, one may reduce the time of beam sweep as compared to the narrowband system, i.e., the beam squint effect surprisingly has the positive effect for sensing the users' positions.

However, the range of the beam squint depends on the transmission bandwidth of the system,
and in general can not meet the requirements of large range coverage.
Hence, we must expand the range of the beam squint, which can be achieved by adopting TDs.

\section{User Position Sensing Scheme with TDs}

We consider that the $N$ phase shifters each cascade on a TD line \cite{b1}.
Assume that the $n$-th phase shifter response is $e^{-j2\pi\phi_n}$.
The corresponding $n$-th TD time domain response is $\delta(t-t_n)$, 
and then its frequency domain response is $e^{-j2\pi \widetilde{f}t_n}$.
The new array beamforming vector assisted by TDs can be expressed as
\begin{equation}
\begin{split}
\begin{aligned}
\label{deqn_ex1a}
[\mathbf{\tilde{w}}]_n=e^{-j2\pi\phi_n}e^{-j2\pi \widetilde{f}t_n}
\end{aligned}
\end{split}
\end{equation}
where $\phi_n$ denotes the phase shift amount of the $n$-th PS, 
$t_n$ denotes the time delay amount of the $n$-th TD,
$\delta(t)$ denotes impulse function, and
$\widetilde{f}$ is the baseband frequency.

With TDs, the power of the $k$-th user at sub-carrier $f_m$ can be expressed as
\begin{equation}
\begin{split}
\begin{aligned}
\label{deqn_ex1a}
&\tilde{g}_{k,m}=\left| \mathbf{h}^H(x_k,y_k,{f}_m)\cdot\mathbf{\tilde{w}}\right|\\
&=\sqrt{\frac{\beta(f_m)}{N}}\left| \sum _{n=-\frac{N-1}{2}}^{\frac{N-1}{2}} 
\frac{1}{r_{k,n}}e^{j2\pi f_m\frac{r_{k,n}}{c}}e^{-j2\pi\phi_n}e^{-j2\pi \widetilde{f}_mt_n}
\right|.
\end{aligned}
\end{split}
\end{equation}


\subsection{Controlled Beam Squint Points Trajectory Based on TDs}

We focus only on the part on the phase in (13). Similarly, $\tilde{g}_{k,m}$ will be maximized when the following equation is satisfied

\begin{equation}
\begin{split}
\begin{aligned}
\label{deqn_ex1a}
f_m\frac{r_{k,n}}{c}-\phi_n-\widetilde{f}_mt_n=p_n.
\end{aligned}
\end{split}
\end{equation}

Although the start and end of the beam squint trajectory are determined by subcarrier $f_0$ and $f_M$.
With the help of TDs, 
we can control the start point and the end point of the beam squint trajectory.

By setting the PSs, 
we can let the beamforming at $f_0$ point to the start point $(r_0,\theta_0)$.
To do so,
we bring ${f}_m={f}_0$, $\widetilde{f}_m=\widetilde{f}_0=0$, $r_k=r_0$, $p_n=0$ into (14), and derive $\phi_{n}=\frac{f_0r_{0,n}}{c}$.
Then we set the phase shift of PSs by $\phi_{n}$.

By setting the TDs, 
we can let the beamforming at $f_M$ point to the end point $(r_c,\theta_c)$.
To do so,
we bring ${f}_m={f}_M$, $\widetilde{f}_m=\widetilde{f}_M=W$, $r_k=r_c$, $p_n=0$ into (14), and derive
$t_{n}=\frac{f_M}{Wc}r_{c,n}-\frac{\phi_{n}}{W}$. Then we set the TDs by $t_{n}$.

As the subcarrier frequency $f_m$ increases, the beam is squint from the start point $(r_0,\theta_0)$ to the end point $(r_c,\theta_c)$. 
For the subcarrier $f_m$, its beam squint point $(r_m,\theta_m)$ still satisfies the equation (14). And we next compute the value of $(r_m,\theta_m)$.

Let us bring $r_k=r_m$, $\phi_{n}=\frac{f_0r_{0,n}}{c}$, $t_{n}=\frac{f_M}{Wc}r_{c,n}-\frac{\phi_{n}}{W}$  into (14). In addition, both ${r}_{m,n}$, $r_{0,n}$ and $r_{c,n}$ are approximated by the Fresnel approximation (5).
Then we obtain

\begin{equation}
\begin{split}
\begin{aligned}
\label{deqn_ex1a}
&\left[\frac{f_m}{c}\left({r}_m+\frac{n^2d^2\cos^2{\theta}_m}{2{r}_m}\right)
-\frac{(W-\widetilde{f}_{m})f_0}{Wc}\right.\\&\left.\left(r_{0}+\frac{n^2d^2\cos^2\theta_{0}}{2r_{0}}\right)
-\frac{f_M\widetilde{f}_{m}}{Wc}\left(r_{c}+\frac{n^2d^2\cos^2\theta_{c}}{2r_{c}}\right)\right]-
\\&\left[\frac{ f_m}{c}nd\sin{\theta}_m-
\frac{(W-\widetilde{f}_{m})f_0}{Wc}nd\sin\theta_{0}-\frac{f_M\widetilde{f}_{m}}{Wc}nd\sin\theta_{c}\right]=p_{n}.
\end{aligned}
\end{split}
\end{equation}

Similar to the derivation of (6) in section \RNum{3}, 
we let the last term of (15) equal to zero, i.e.
$nd\left[\frac{f_m}{c}\sin{\theta}_m-\frac{(W-\widetilde{f}_{m})f_0}{Wc}\sin\theta_{0}-\frac{f_M\widetilde{f}_{m}}{Wc}\sin\theta_{c}\right]=0$. 
Then the angle ${\theta}_m$ satisfies
\begin{equation}
\begin{split}
\begin{aligned}
\label{deqn_ex1a}
\sin{\theta}_m=\frac{(W-\widetilde{f}_{m})f_0}{Wf_m}\sin\theta_{0}+\frac{(W+f_0)\widetilde{f}_{m}}{Wf_m}\sin\theta_{c}.
\end{aligned}
\end{split}
\end{equation}

Then, (15) reduces to
\begin{equation}
\begin{split}
\begin{aligned}
\label{deqn_ex1a}
&\frac{f_m}{c}\left({r}_m+\frac{n^2d^2\cos^2{\theta}_m}{2{r}_m}\right)
-\frac{(W-\widetilde{f}_{m})f_0}{Wc}\left(r_{0}+\right.\\&\left.\frac{n^2d^2\cos^2{\theta}_0}{2r_{0}}\right)
-\frac{f_M\widetilde{f}_{m}}{Wc}(r_{c}+\frac{n^2d^2\cos^2{\theta}_c}{2r_{c}})
=p_{n}.
\end{aligned}
\end{split}
\end{equation}

For $n=0$, we obtain from (17) that
\begin{equation}
\begin{split}
\begin{aligned}
\label{deqn_ex1a}
&\frac{f_m}{c}{r}_m
-\frac{(W-\widetilde{f}_{m})f_0}{Wc}r_{0}
-\frac{f_M\widetilde{f}_{m}}{Wc}r_{c}
=p_{0}.
\end{aligned}
\end{split}
\end{equation}

We then bring (18) into (17) and obtain

\begin{equation}
\begin{split}
\begin{aligned}
\label{deqn_ex1a}
\frac{d^2}{2c}&\left[
\frac{f_m}{{r}_m}\cos^2{\theta}_m-\frac{(W-\widetilde{f}_{m})f_0}{Wr_{0}}\cos^2{\theta}_0\right.\\&\left. -\frac{f_M\widetilde{f}_{m}}{Wr_{c}}\cos^2{\theta}_c
\right]=\frac{p_{n}-p_{0}}{n^2}.
\end{aligned}
\end{split}
\end{equation}

Considering the arbitrariness of $n$ in (19), we know that
$\frac{f_m}{{r}_m}\cos^2{\theta}_m-\frac{(W-\widetilde{f}_{m})f_0}{Wr_{0}}\cos^2{\theta}_0-\frac{f_M\widetilde{f}_{m}}{Wr_{c}}\cos^2{\theta}_c=0$. Then the distance ${r}_m$ satisfy

\begin{equation}
\begin{split}
\begin{aligned}
\label{deqn_ex1a}
\frac{1}{{r}_m}=\frac{1}{r_{0}}\frac{(W-\widetilde{f}_{m})f_0}{Wf_m}\frac{\cos^2{\theta}_0}{\cos^2{\theta}_m}+
\frac{1}{r_{c}}\frac{(W+f_0)\widetilde{f}_{m}}{Wf_m}\frac{\cos^2{\theta}_c}{\cos^2{\theta}_m}
\end{aligned}
\end{split}
\end{equation}
where ${\theta}_m$ is given by equation (16).

\subsection{User Angle Sensing Scheme}

Suppose the sensing range required by the system (where all $K$ users stay) is a sector region that the angle changes from $\theta_{max}$ to $\theta_{min}$ and the distance changes from $r_{min}$ to $r_{max}$.
Assume that $r_{mid}\in[r_{min},r_{max}]$ is an appropriate specific distance value.

We target to use one time beam sweep to get all $K$ users' angles $\hat{\theta}_k$.
The specific steps are as follows:

\begin{enumerate}

\item Let the beamforming from $f_0$ point to the start point 
$(r_{mid},\theta_{max})$ by setting the PSs.

\item Let the beamforming from $f_M$ point to the end point 
$(r_{mid},\theta_{min})$ by 
setting the TDs.

\item BS simultaneously transmits $M$ sub-carriers, 
the angles change from $\theta_{max}$ to $\theta_{min}$, 
and the distances change from $r_{mid}$ to $r_{mid}$, 
covering the whole space in the form of a curve.

\item  All users simultaneously receive $M$ sub-carriers and feed back the maximum power sub-carrier frequencies $f_{1,d1}$, $f_{2,d1}$,..., $f_{K,d1}$ to the BS. 
Then BS can calculate the angle of the $k$-th user by substituting 
$\theta_m=\hat{\theta}_k$, $f_m=f_{k,d1}$ and $\widetilde{f}_m=f_{k,d1}-f_0$ into (16).

\end{enumerate}

It should be noted that if more than two users are in the same angle direction, their angle estimates will be the same. 
This means that the beam sweep will get $\check{K}$ different angle estimates,
where $\check{K}\leq K$ and $\check{K}\leq M$. Let us mark this $\check{K}$ different angles as 
$\check{\theta}_1,\check{\theta}_2,...,\check{\theta}_{\check{K}}$.

\subsection{User Distance Sensing Scheme}

Assume that $\check{\theta}_q=\hat{\theta}_k$ and 
there are $K_q$ users in the $\check{\theta}_q$ direction.
We next sense the user-$k$'s distance $\hat{r}_k$ in the $q$-th time
beam sweep with the following steps:

\begin{enumerate}

\item Let the beamforming from $f_0$ point to the start point $(r_{min},\hat{\theta}_k)$ by setting the PSs.

\item Let the beamforming from $f_M$ point to the end point $(r_{max},\hat{\theta}_k)$ by setting the TDs.

\item BS simultaneously transmits $M$ sub-carriers
and all the angles are $\hat{\theta_k}$ ($=\check{\theta}_q$),
but the distances change from $r_{min}$ to $r_{max}$, 
covering the whole space in the form of a straight segment.

\item  The $K_q$ users simultaneously receive $M$ sub-carriers and feed back the maximum power sub-carrier frequencies $f_{q1,d2}$, $f_{q2,d2}$,..., $f_{qK_q,d2}$ to the BS. 
Then BS can calculate the distance of the $k$-th user by substituting 
$\theta_m=\hat{\theta}_k$, $r_m=\hat{r}_k$, $f_m=f_{k,d2}$ and $\widetilde{f}_m=f_{k,d2}-f_0$ into (20).

\end{enumerate}

The above steps for distance sensing will be repeated $\check{K}$ times to obtain the distance of each group of users in different directions.
Then we will obtain all $K$ users' position estimates
$(\hat{r}_1,\hat{\theta}_1),(\hat{r}_2,\hat{\theta}_2),...,(\hat{r}_K,\hat{\theta}_K)$.

Remark: The beam sweeping to sense all $K$ users' angles is required only once, while the beam sweeeping to sense all $K$ users' distances, in the worst case, are required $M$ times.

\begin{figure}[!t]
\centering
\includegraphics[width=3in]{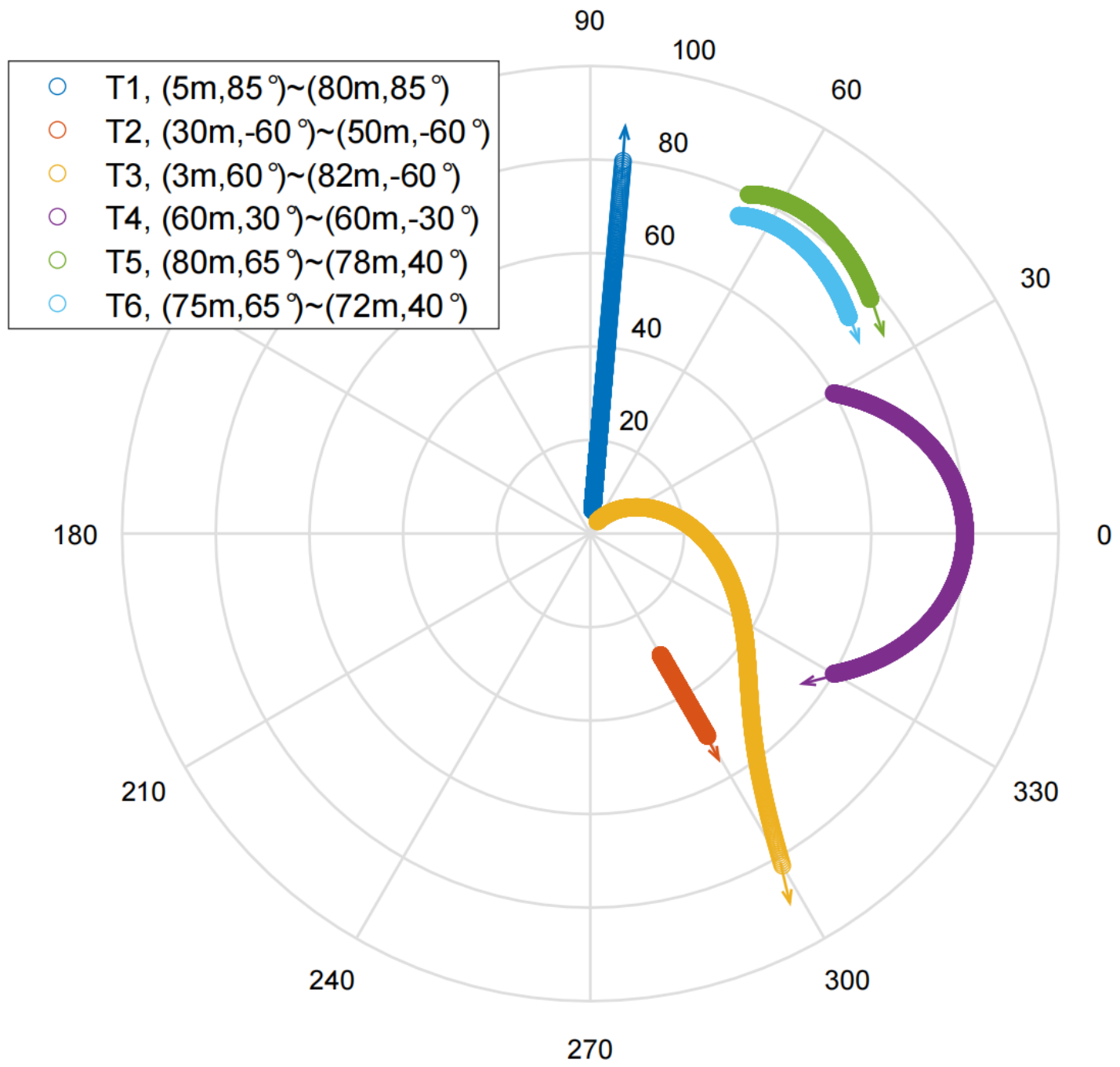}
\caption{TDs-assisted near-field beam squint trajectories.}
\label{fig2:env}
\end{figure}

\begin{figure}[!t]
\centering
\includegraphics[width=3.5in]{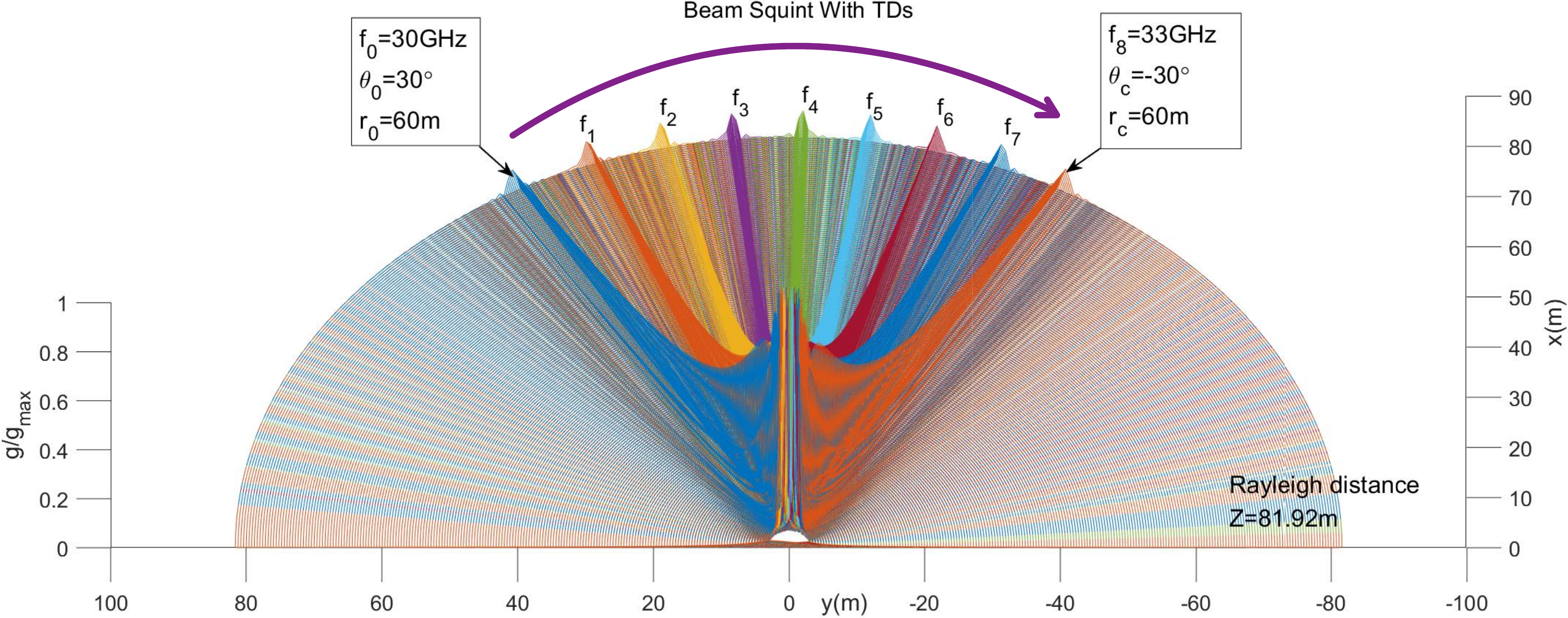}
\caption{Trajectory of beam squint. Where $f_0=$ 30 GHz, $W=$ 3 GHz, $M=8$, the lowest frequency carrier $f_0$ is set to focus on $(60m, 30^{\circ}).$}
\label{fig_1}
\end{figure}

\section{Simulation Result }

In simulation, we consider the number of antennas for BS as $N=128$, the lowest carrier frequency $f_0$ = 30 GHz and OFDM with bandwidth $W$ = 3 GHz.
Thus the central carrier wavelength is $\lambda=0.01m$, and the antenna spacing is $d=\frac{1}{2}\lambda$.
Then the Rayleigh distance can be computed as $Z=81.92 m$. 
The noise is assumed to obey the complex Gaussian distribution with mean $\mu=0$ and  variance $\sigma^2=1$.

\subsection{TDs-assisted Near-field Beam Squint}

With the aid of TDs, we can control the start point and the end point of the beam squint trajectory, and some examples of different controlling are shown in Fig. 3, where the number of subcarriers is assumed to be $M=2048$.

Trajectory T1 and trajectory T2 are two straight lines in the radial directions, but with different distance range sizes, where T1 squint from $(5m,85^\circ)$ to $(80m,85^\circ)$ while T2 squint from $(30m,-60^\circ)$ to $(50m,-60^\circ)$. The range of the time delay of TDs required are $t_{n,T1}\in[2.7656\mu s,2.7677\mu s]$ and $t_{n,T2}\in[0.8324\mu s,0.8342\mu s]$, respectively.
If we already know the user's angle, using the radius trajectories like T1 or T2 allows to quickly locate the user's distance.
Trajectory T3 squints from $(3m,60^\circ)$ to $(82m,-60^\circ)$, and the range of squint spans the entire expected sensing angle and distance regions, for which the TDs required is $t_{n,T3}\in[2.8873\mu s,2.9258\mu s]$.
Trajectory T4 squints from $(60m,30^\circ)$ to $(60m,-30^\circ)$ with
$t_{n,T4}\in[0.1889\mu s,0.2112\mu s]$ in a symmetrical form. This symmetry is more conducive to obtaining the user's angle.
Trajectories T5 and T6 show beam squint in a smaller range, which favors the coverage of already-known narrow regions.

\begin{figure}[!t]
\centering
\includegraphics[width=3.4in]{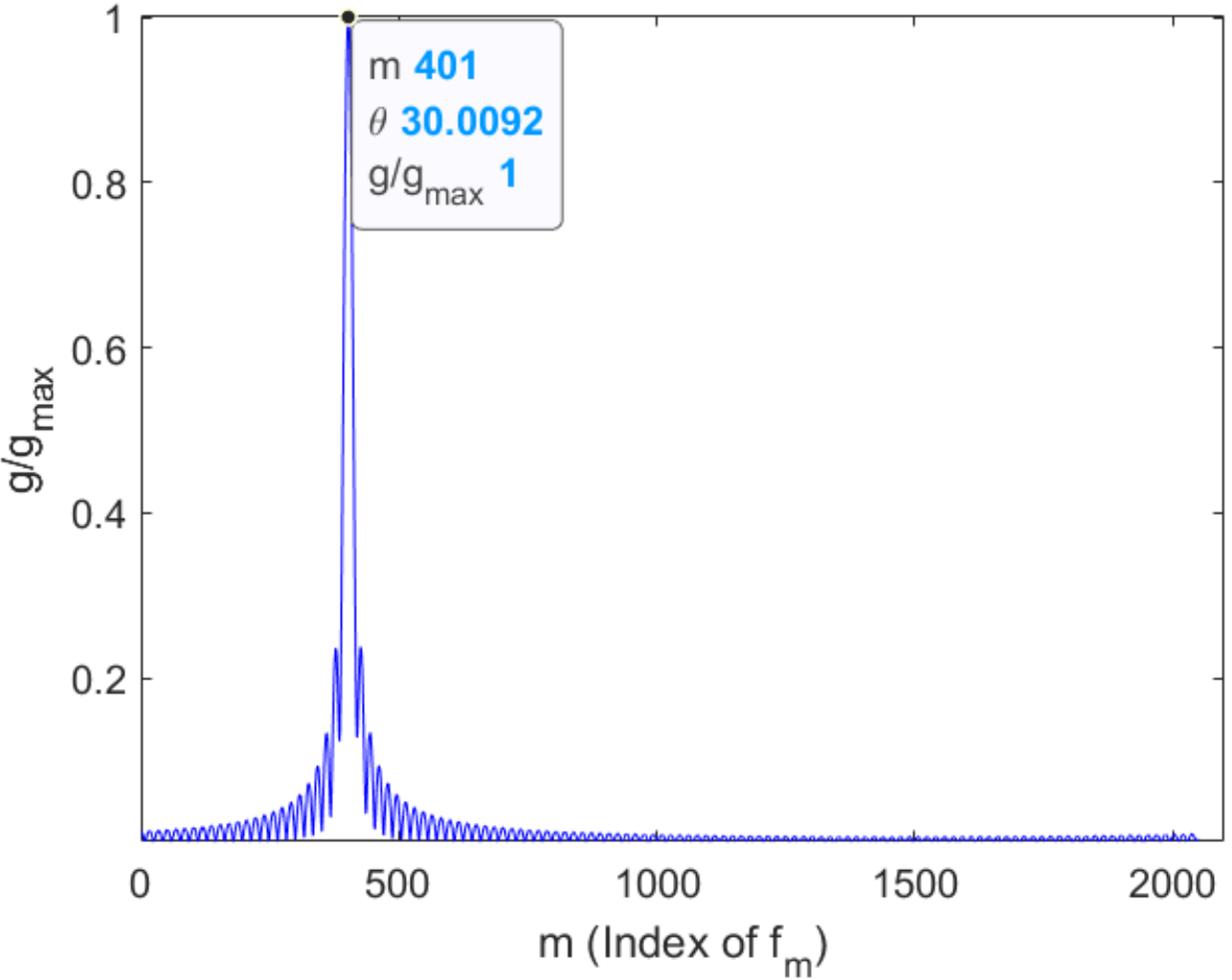}
\caption{Normalized power gain for the different subcarriers received by the user at $(30m,30^\circ)$ position in angle sensing stage.}
\label{fig_1}
\end{figure}

\begin{figure}[!t]
\centering
\includegraphics[width=3.4in]{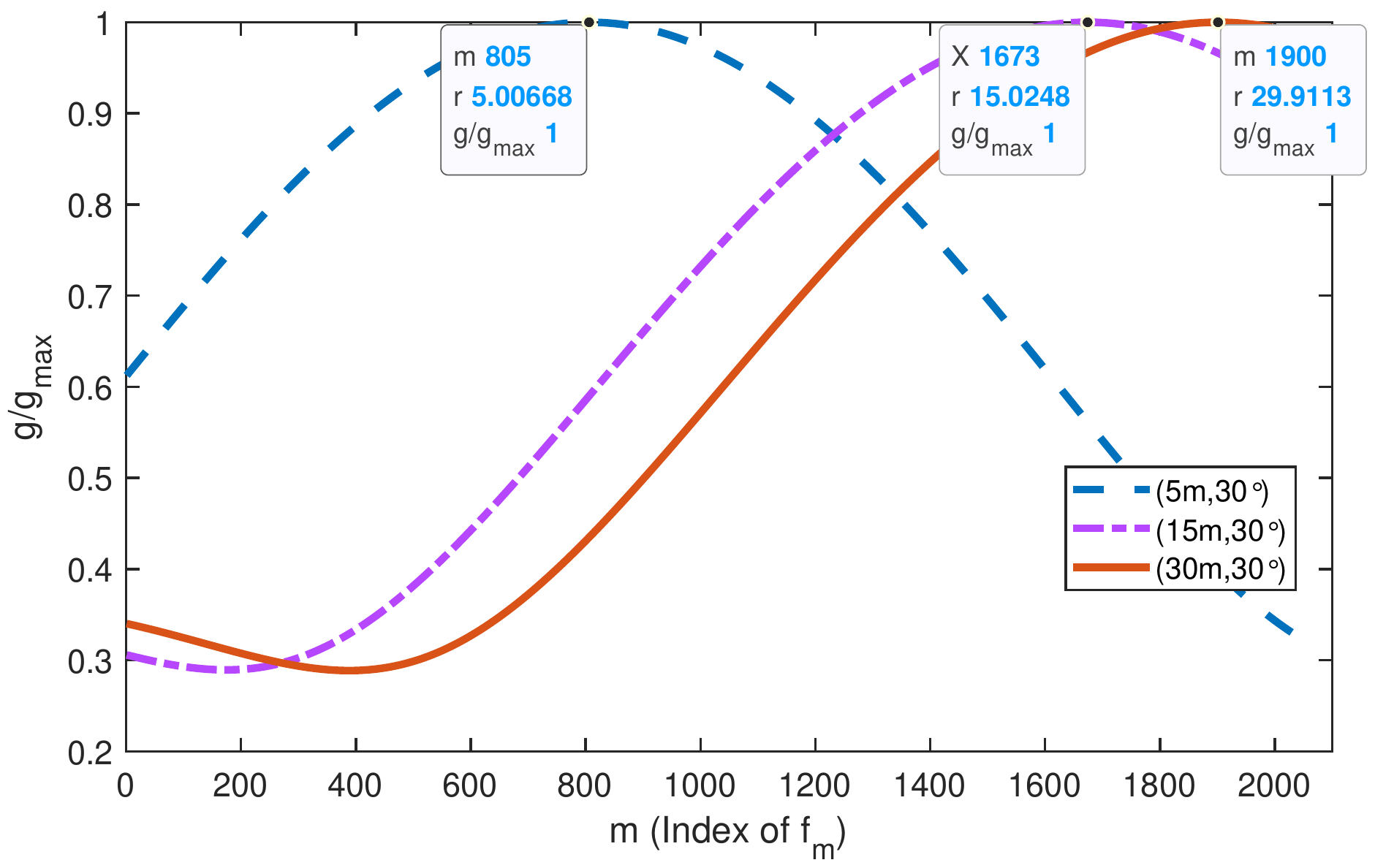}
\caption{Normalized power gain for the different subcarriers received by the users at $(5m,30^\circ)$, $(15m,30^\circ)$ and $(30m,30^\circ)$ position in distance sensing stage.}
\label{fig_1}
\end{figure}

Let us now show how the trajectory T4 can obtain all users' angles.
We let the $M=8$ to clearly show the beamforming from different subcarriers.
We can calculate the gain of each subcarrier at each point by formula (13).
Fig. 4 shows the normalized beamforming gain of all discrete field points 
through the Electromagnetic distance $3.17m$ to Rayleigh distance $81.92m$ by distance step $0.4m$ and $90^\circ$ to $-90^\circ$ by angle step $-0.5^\circ$.
As the frequency gradually increases, the subcarrier squints from $30^\circ$ to $-30^\circ$, and different subcarriers have peaks in different directions. With this feature, we can determine the user's angle according to the subcarrier frequency of the maximum power feedback by the user, which will be densely effective when the number of subcarriers $M$ is large enough.

\begin{figure*}[!t]
\centering
\subfloat[]{\includegraphics[width=3in]{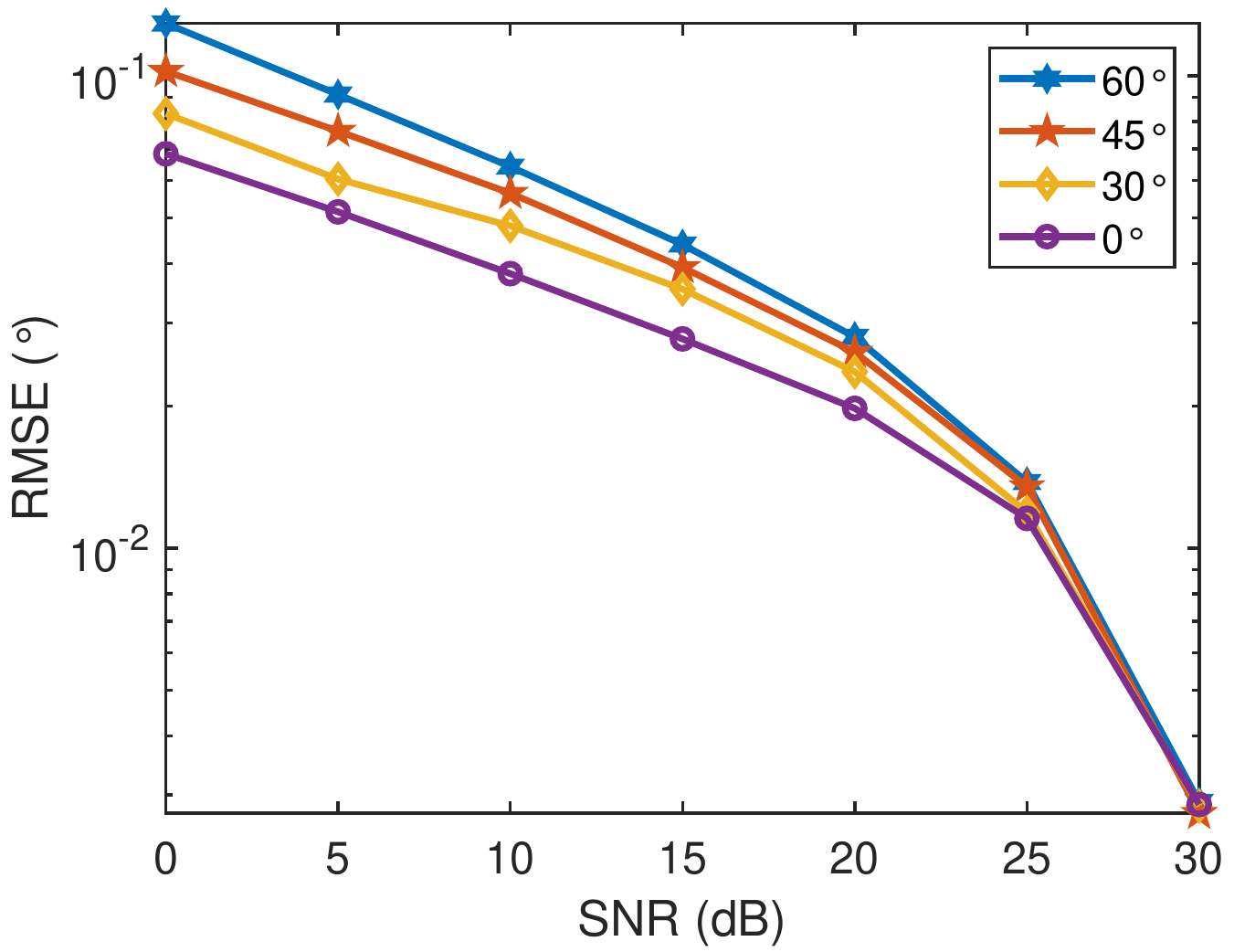}%
\label{fig_first_case}}
\hfil
\subfloat[]{\includegraphics[width=3in]{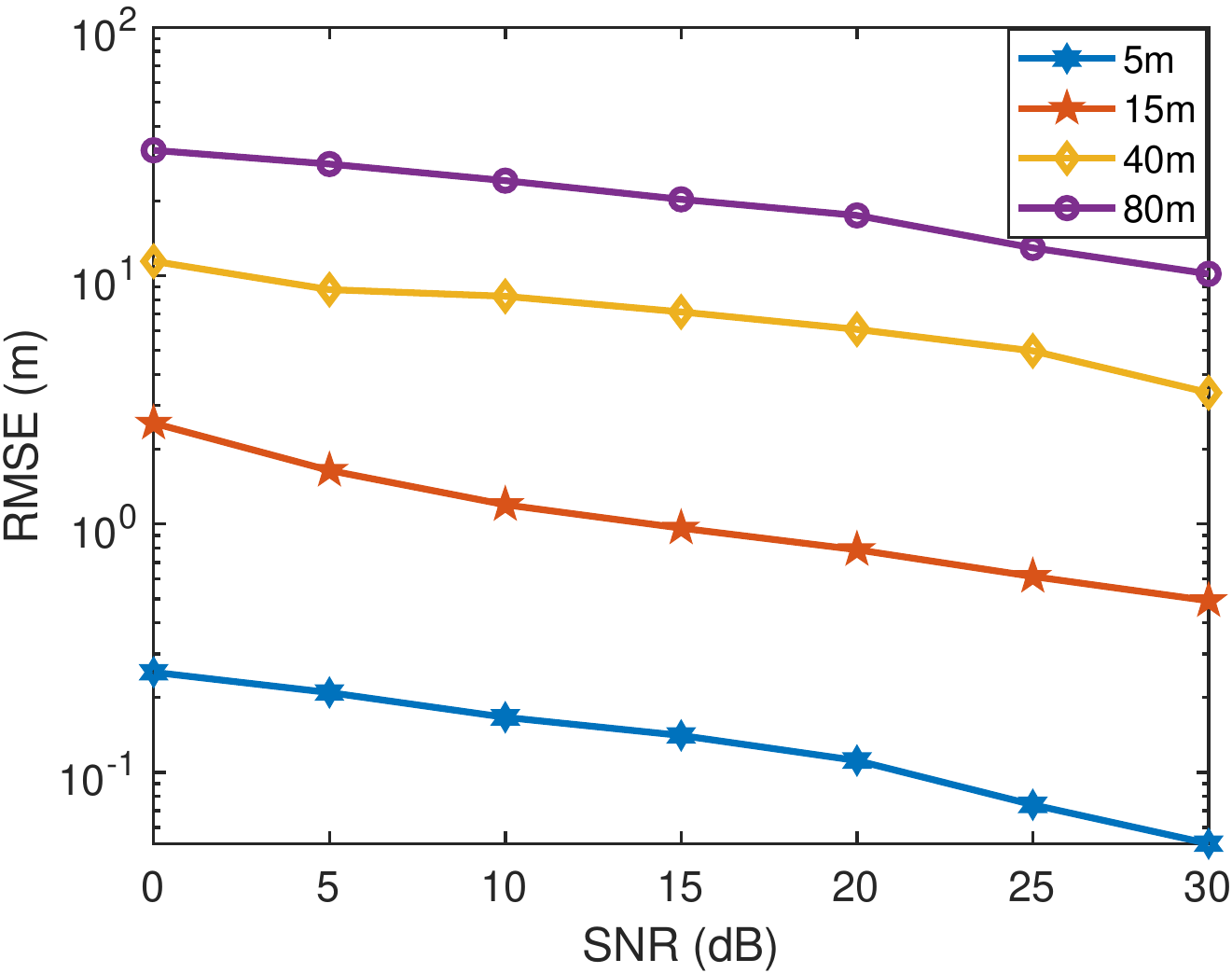}%
\label{fig_second_case}}
\caption{(a) The RMSE of user-angle-sensing, where user's distance is $40m$.
(b) The RMSE of user-distance-sensing, where user's angle is $0^\circ$.}
\label{fig_sim}
\end{figure*}

\subsection{User Angle Sensing Scheme With TDs}

Consider the Electromagnetic and Rayleigh distances, 
and set the sensing range of BS system is $[\theta_{max},\theta_{min}]=[60^{\circ},-60^{\circ}]$,
$[r_{min},r_{max}]=[3m,82m]$, and the number of subcarriers $M=2048$.
Assume that there is a user located at $(30m,30^\circ)$.

Following the proposed scheme,
we set $(r_{mid}, \theta_{max})=(40m,60^{\circ})$ and $(r_{mid}, \theta_{min})=(40m,-60^{\circ})$,
and then the beam squint trajectory covers the entire angle range similar to T4.
The user will receive all subcarriers with different subcarrier power as shown in Fig. 5. 
We can get the $401$-th subcarrier has the maximum power, and then calculate the user's angle as $\hat{\theta}=30.0092^\circ$ from $f_{401}$.

\subsection{User Distance Sensing Scheme With TDs}

After obtaining the user's angle, let us obtain the user's distance.
Following the proposed scheme,
we set $(r_{min}, \hat{\theta})=(3m,30.0092^{\circ})$ and $(r_{mid}, \hat{\theta})=(82m,30.0092^{\circ})$,
and then the beam squint trajectory covers the entire distance range similar to T1.
The user in Fig. 5 will receive all subcarriers with different subcarrier power as shown in Fig. 6.
We can get the $1900$-th subcarrier has the maximum power,
and then calculate the user's distance as $\hat{r}=29.9113m$ from $f_{1900}$.

Furthermore, the normalized power of the different subcarriers received by another two users in the $30^\circ$ direction is given in Fig. 6.
By comparing the three curves, we can see that the users at different distances receive the maximum power at different subcarriers, which allows us to locate the user's distance.
Note that the power change of different subcarriers near the maximum power is flat, which is very different from the angle sensing.
This means that distance estimation is easily subject to the noise, and distance estimation is not as accurate as angle estimation, and we may need to consider using frequency conversion or multiple BSs methods to improve this problem.
Moreover, Fig. 6 shows the inhomogeneity of the beam squint, with a large number of subcarriers clustered on the side near the start point, while the side near the end point is too small. This phenomenon leads to large distance localization error at slightly farther locations when $M$ is limited.
In this case, we can use a rough scan to obtain the user's approximate position, and then make a small range of beam squint near this position, in order to obtain a more accurate position of the user.

\subsection{Multi-Users Localization under Different SNR}

Assuming that multiple users are randomly and evenly distributed within the BS sensing range.
Fig.7 shows the root-mean-square error (RMSE) of the users’ positions sensing.

Fig.7 (a) shows that when the user's distance is fixed, the angle-sensing RMSE keeps decreases as SNR increases.
The average angle sensing accuracy is at $0.1^\circ$ when the SNR is $0$ dB, and reaches $0.01^\circ$ when the SNR is increased over $25$ dB. In addition, the closer the user's angle is to $0^\circ$, the lower the angle-sensing RMSE, and the higher the angle-sensing accuracy. The gap is more pronounced when the SNR is low.

Fig.7 (b) shows that when the user's angle is fixed, as the SNR increases, the distance-sensing RMSE is also decreasing. When the user is closer to the BS, the distance sensing accuracy is higher.
For the users within $15m$, the localization accuracy can reache at $1m$ when the SNR exceeds $10$ dB.
For users within $40m$, the localization accuracy exceeds $3m$ when the SNR exceeds $30$ dB.
However, for the users near the Rayleigh distance, the localization RMSE is large when the SNR is low, which means that the applicability of the near-field model near the Rayleigh distance is weakened.

\section{Conclusion}

We propose a fast user localization scheme based on beam squint in near-field communications system.
We first find that with the aid of the TDs, the range and trajectory of the beam squint can be freely controlled, and hence it is possible to reversely utilize the beam squint for user localization.
Then we derive the trajectory equation for near-field beam squint points and design a way to control the trajectory of these beam squint points.
With the proposed design, beamforming from different subcarriers would purposely point to different angles and different distances such that users from different positions would receive the maximum power at different subcarriers. Hence, one can simply find the different users' position from the beam squint effect.  Simulation results demonstrate the effectiveness of the proposed scheme.

\bibliographystyle{ieeetr}
\bibliography{refer.bib}

\vfill

\end{document}